\documentstyle[12pt,epsf,aps]{revtex} 

\newcommand{\bfk}{{\bf k}}
\newcommand{\bfx}{{\bf x}}

\newcommand{\bge}{\begin{equation}}
\newcommand{\ee}{\end{equation}}

\newcommand{\hcc}{H_{\mbox{\scriptsize cc}}}
\newcommand{\hcd}{H_{\mbox{\scriptsize cd}}}
\newcommand{\hdc}{H_{\mbox{\scriptsize dc}}}
\newcommand{\hdd}{H_{\mbox{\scriptsize dd}}}

\renewcommand{\baselinestretch}{1.5}
\begin{document}

\title{
{\bf Shear-Induced Isotropic-to-Lamellar Transition in a 
Lattice-Gas Model of Ternary Amphiphilic Fluids}
}
\author{
Andrew N. Emerton \footnote{Present address: Smith System Engineering
Ltd., Guildford Research Park, Guildford Surrey GU2 5YP, U.K.} \\
{\small \sl Department of Theoretical Physics, Oxford University,}\\
{\small \sl 1 Keble Road, Oxford OX1 3NP, U.K.}\\
Florian W.J. Weig \footnote{Present address:
Ludwig-Maximilians-Universit\"{a}t, Theoretische Physik, Theresienstra\ss
e 37, D-80333 M\"{u}nchen, Germany}\\
{\small \sl Department of Theoretical Physics, Oxford University,}\\
{\small \sl 1 Keble Road, Oxford OX1 3NP, U.K.}\\
Peter V. Coveney \footnote{Author to whom correspondence should be
addressed }\\
{\small \sl Schlumberger Cambridge Research,}\\
{\small \sl High Cross, Madingley Road, Cambridge CB3 0EL, UK.}\\
{\small \sl and Department of Theoretical Physics, Oxford University,}\\
{\small \sl 1 Keble Road, Oxford OX1 3NP, U.K.}\\
Bruce M. Boghosian\\
{\small \sl Center for Computational Science, Boston University,}\\
{\small \sl 3 Cummington Street, Boston, Massachusetts 02215, U.S.A.} \\
[0.3cm]
}
\date{\today}
\maketitle

\begin{abstract}
Although shear-induced isotropic-to-lamellar transitions in ternary
systems of oil,
water and surfactant have been observed experimentally and predicted
theoretically by simple models for some time now, their numerical
simulation has not been achieved so far. In this work we demonstrate
that a recently introduced hydrodynamic lattice-gas model of amphiphilic
fluids is well suited for this purpose: the two-dimensional
version of this model does indeed exhibit a
shear-induced isotropic-to-lamellar phase transition.

\noindent PACS numbers: 82.70.-y;05.70.Lm
\end{abstract}

\section{Introduction}

Soft materials such as polymer solutions, liquid crystals,
surfactants, and microemulsions are frequently processed or utilized
through the application of large deformations. Various attempts have
been made to investigate and characterise the behaviour of such
complex systems under conditions such as shear flow, for which there
are numerous industrial applications \cite{bib:h}. In this letter we   
model the effect of linear shear flow on a hydrodynamic,
isotropic, sponge microemulsion phase using our recently 
introduced lattice-gas automaton model for simulating self-assembling
amphiphilic systems \cite{bib:bce}. Experimental results
\cite{bib:ktb,bib:yt,bib:mmk} as well as theoretical predictions
\cite{bib:cm} provide
evidence for the presence of a transition from an isotropic to a
lamellar phase in such systems; however, we are unaware of any model
that is capable of {\it simulating} complex
multi-phase flow of this sort. This is because of the considerable
difficulty involved in simulating ternary amphiphilic systems under
hydrodynamic flow, as well as in the implementation of the 
boundary conditions required for shear. Traditional continuum based
fluid-dynamical modelling methods, such as finite-difference,
finite-element or
volume-of-fluid techniques cannot viably deal with the complexity
involved; molecular dynamics, on the other hand, are too
computationally expensive.  

Hydrodynamic lattice-gas models have evolved from simulating simple 
one-component Navier-Stokes fluids \cite{bib:fhp}, to two- and 
multi-component immiscible fluids \cite{bib:gr}. We have
recently extended such models to systems including amphiphiles  
\cite{bib:bce}. Lattice-gas models can reproduce
fluid dynamics on mesoscopic and higher levels, permitting the investigation of
non-equilibrium (kinetic) behaviour over a broad range of length and
time scales \cite{bib:rz}. The relative simplicity of the collision
rules, the self-assembly of complex interfaces, the presence of
natural underlying kinetic fluctuations and the ease of implementation
of complex boundary conditions suggest that such hydrodynamic 
lattice-gas models are
an appropriate choice for the study of amphiphilic fluids under
linear shear.    

Linear shear flow has previously been applied to binary fluid systems
in both two \cite{bib:r} and three-dimensional \cite{bib:or}
lattice-gas models. We make use of the method introduced in
the second of these papers for obtaining linear shear flow on the
lattice, although some modification is required for our ternary 
amphiphilic system. 

As well as simulating the shear-induced isotropic-to-lamellar
transition, this work
was undertaken in order to investigate further the validity of our
amphiphilic model for complex fluid simulation. This is the first
application of the model to the simulation of known physical effects
associated with bulk fluid flow. In Section \ref{sec:ma} we briefly
describe our model and the numerical techniques we have used to
investigate the system under shear; our simulations are presented in
Section \ref{sec:r} with concluding remarks in Section \ref{sec:conc}. 

\section{Model and Analysis}
\label{sec:ma}

We perform simulations using our hydrodynamic lattice-gas model of 
amphiphilic systems as previously published \cite{bib:bce}. The model
is a microscopic dynamical system which gives the correct mesoscopic
and macroscopic behaviour of mixtures of oil, water and
surfactant. The model is based on the two-fluid immiscible lattice-gas
of Rothman and Keller \cite{bib:rk}, which we have reformulated using a
microscopic particulate description to permit the inclusion of
amphiphile. Pursuing the electrostatic analogy with the Rothman-Keller
model, we describe surfactant
molecules as {\it dipoles}, characterised by a dipole vector {\bf
$\sigma$}. The model exhibits the commonly formed equilibrium
microemulsion phases, including droplets, sponge structures (the
two-dimensional analogue of the bicontinua in three dimensions) and lamellae
\cite{bib:bce}. 
Moreover, the lattice-gas model conserves momentum as well as the
masses of the various species, and correctly simulates fluid dynamical
and scaling behaviour during self-assembly of these phases
\cite{bib:ecb,bib:wcb}.

It should be noted that, formally, no lamellar phase
can exist at finite temperature in two spatial dimensions:
thermal fluctuations are large enough to destroy true long-range
smectic order. 
We showed in our original paper that the stability of such
relatively small and artificially created structures
is greatly enhanced compared to that observed in the absence of 
surfactant or when amphiphilic interactions are
extinguished in a ternary fluid~\cite{bib:bce}. A more thorough
investigation of the relative stability of the observed lamellae
remains as work for the future.   

In order to incorporate the most general form of interaction energy
within our model system, we introduce a set of coupling
constants $\lambda, \mu, \epsilon, \zeta$, in terms of which the total
interaction energy can be written as \cite{bib:bce}
\bge
\Delta H_{\mbox{\scriptsize int}}
       =   \lambda \Delta \hcc +
         \mu \Delta \hcd +
         \epsilon \Delta \hdc +
         \zeta \Delta \hdd.
\label{eq:tiw}
\ee
The four terms on the right hand side correspond, respectively, to the
relative immiscibility of oil and water, the tendency of surfactant to 
surround oil or water droplets, the propensity of surfactant dipoles
to align across oil-water interfaces and the contribution from
pairwise interactions between surfactant molecules. For consistency we
choose these coefficients to be of the same value for all simulations
in this paper, which allow a sponge microemulsion phase to
form in one part of the ternary phase diagram \cite{bib:bce}; as such
these are,  
\bge
 \lambda = 1.0, \mu = 0.05, \epsilon = 8.0, \zeta = 0.5.
 \label{eq:dcc}
\ee

\begin{figure}
\begin{center}
\leavevmode
\hbox{%
\epsfxsize=4.0in
\epsffile{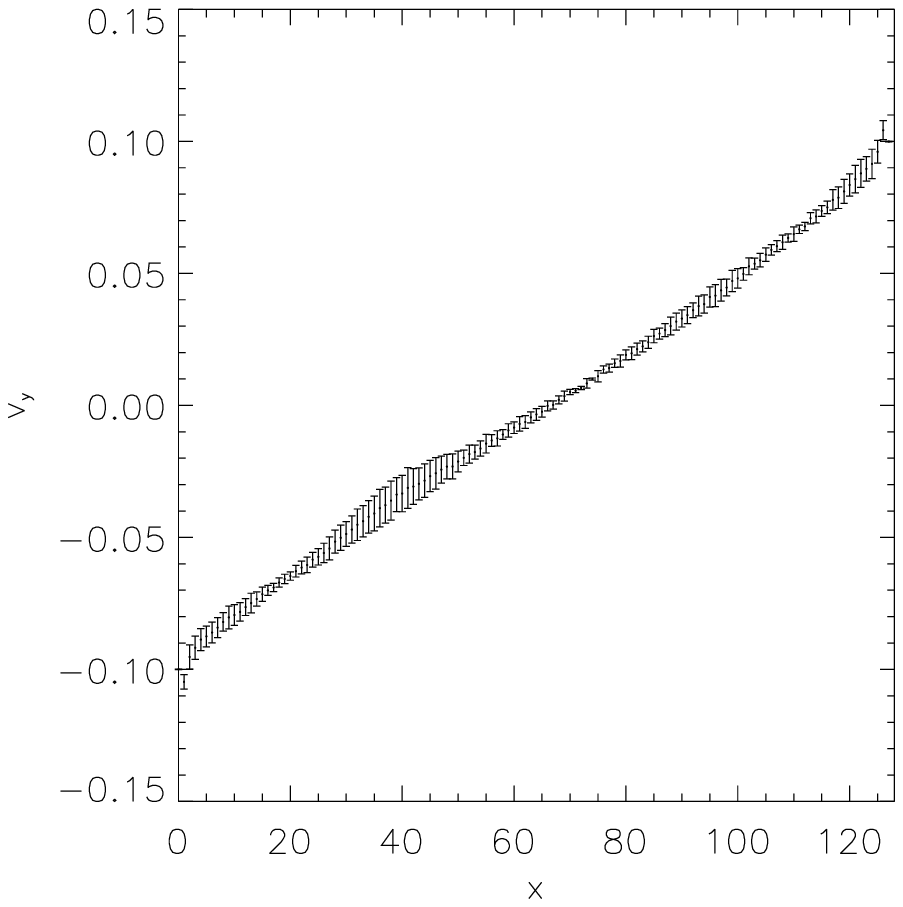}}
\end{center}
\caption{\sl Vertical (y-direction) velocity against column number
(x-direction) for
a $128 \times 128$ lattice. The data points are averaged over $300$
measurements starting at time step $10,000$, when the steady state has
been reached. The error bars result from
ensemble-averaging over five independent runs.}
\label{fig:yvelprof}
\end{figure}

As stated above, in order to investigate the effect of linear shear
flow on a sponge microemulsion phase we apply a technique devised by
Olson and Rothman \cite{bib:or} for shearing two-component binary
lattice-gas fluids, which we have adapted to meet the requirements of
our amphiphilic lattice-gas model. The technique provides a linearly
varying, tunable velocity gradient in a direction orthogonal to the
flow (see Fig.~\ref{fig:yvelprof}). The velocities used are small enough
to ensure that the lattice gas accurately represents the hydrodynamic flow
\cite{bib:or}.

We analyse the simulation results in three ways. The first is 
direct visualisation of the growth of domains both with and
without the imposition of a shear velocity. The second, performed in  
order to observe any anisotropic domain growth and the formation of a
characteristic Bragg peak in the presence of shear,
is a quantitative analysis of the structure  
factor for the oil-water density difference,  
\bge
S(\bfk, t) = \frac{1}{N}\left|\sum_{\bfx} (q(\bfx,t) - q^{av})
                   e^{i\bfk\cdot\bfx}\right|^2 ,
\ee
where $\bfk = \left(2 \pi / L \right) \left( m {\bf i} + n {\bf j}
\right)$, $m,n = 1,2,...,\frac{L}{2}$, q(\bfx,t) is the water-minus-oil order
parameter at site ${\bf x}$ and time step $t$, $q^{av}$ is the average
value of this order parameter, $L$ is the length of the system and $N
= L^2$ is the number of lattice sites in the system.
In order to improve statistics and to reduce fluctuation effects we
calculate the running time-average of this structure factor over $T$
measurements at times $t_i =
\tau + A \frac{i}{T}$, where $i$ runs from $1$ to $T$ and $A$ is the
number of time steps after which we start a new running average. In our
simulations we choose $A = 4,000$ and $T = 100$. The time-averaged
structure factor is then given as 
\bge
\bar{S}(\bfk, \tau) = \frac{1}{T} \sum_{i=1}^{T} S(\bfk, \tau +
A \frac{i}{T})
\ee    
Thirdly, we evaluate the values of the
average $x$ and $y$ components of all the surfactant dipole vectors in the
system at each time $t$, namely:
\bge 
X^2 = \sum_\bfx \left( \sum_i \sigma_{ix}^2(\bfx) \right), \label{X}
\ee
and,
\bge
Y^2 = \sum_\bfx \left( \sum_i \sigma_{iy}^2({\bfx}) \right), \label{Y}
\ee
where $\sigma_{ix}$ is the $x$ component of the surfactant dipole 
vector moving in direction $i$ at lattice site $\bfx$, and similarly 
$\sigma_{iy}$ is the $y$ component of that vector. A significant difference
between $X^2$ and $Y^2$ not only indicates anisotropic ordering in the
system, but also provides evidence for the formation of aligned surfactant
layers between oil and water interfaces, which are typical for the
lamellar phase.   

As well as undertaking simulations both with and without a 
shear velocity present for the entire duration of the run, we also 
investigate the case where the imposed velocity becomes non-zero only 
after a predetermined number of time steps of the simulation, 
in order to verify that the transition to the lamellar phase
can indeed be accessed from a perturbation of the equilibrium
isotropic sponge phase (as Cates and Milner assumed
\cite{bib:cm}) and is not just a shear-induced pattern of
phase-separated fluid domains resulting from the initial
configuration, which is a random mixture of the three fluids
\cite{bib:c}.   

\section{Simulations}
\label{sec:r}

As already discussed, we perform simulations both with and without 
shear flow present in order for a critical comparison to be made; in
the former case the shear velocity is introduced at two different
stages during the simulations, either at time step $0$ or after time step 
$10,000$. For all simulations reported here
we use reduced densities for water, surfactant and oil of $0.215$,
$0.1075$ and $0.215$ respectively, although we note that these are
just one set of many in their vicinity that give the
behaviour we describe below. 
We use a $2D$ lattice of size $N \times N$, where $N$ takes the values
$64, 128$ and $256$, with periodic boundary conditions in both dimensions.  
With varying system size we also have to change the shear velocity imposed
on the left and right ($y$ direction) sides of the simulation box in
order to produce the same velocity gradient. For $N = 128$ we
choose velocities of $-0.1$ and $+0.1$ lattice units per time step for
the left and right side. Therefore we use $-0.05$ and $+0.05$
as  velocities
for $N = 64$ and $-0.2$ and $+0.2$ for $N=256$. In each case the
initial condition is a random configuration of all three particle 
types in the system.    
The actual performance of the simulations proved to be
computationally very intensive, especially for the larger system
sizes ($N = 256$), where a typical run computing $40,000$ time steps took 2.5
days on a Sparc Ultra Enterprise 3000.   

\subsection{Absence of Shear}

\begin{figure}
\begin{center}
\leavevmode
\hbox{%
\epsfxsize=5.5in
\epsffile{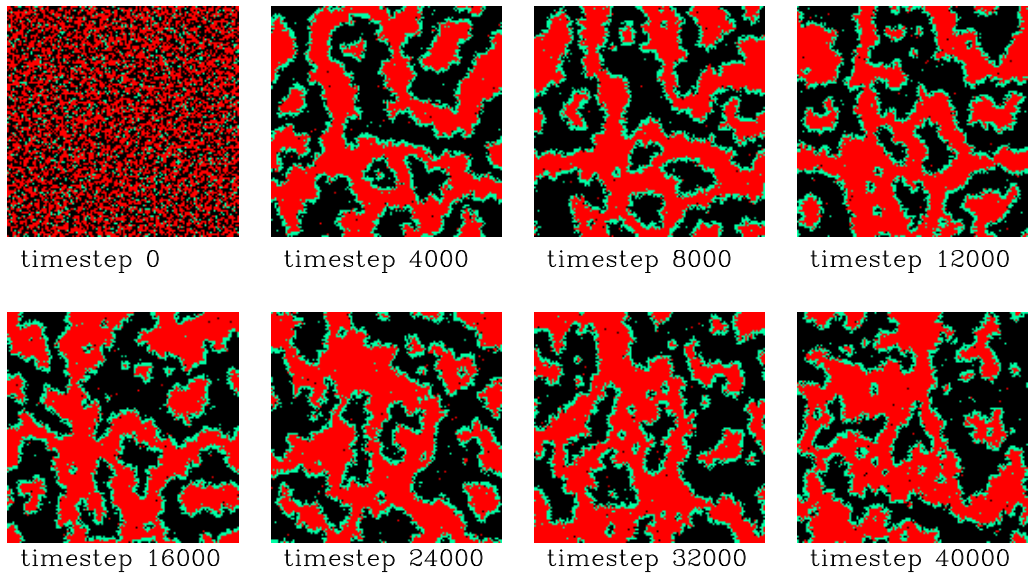}}
\end{center}
\caption{\sl Time evolution of sponge microemulsion phase in 
absence of shear. The $x$ axis is along the horizontal and the $y$ 
axis along the vertical side of the simulation images shown here. The
system size is $128 \times 128$.}
\label{fig:noshearvis}
\end{figure}

The first set of simulations are with
zero shear velocity. We performed five independent simulations over
$40,000$ time steps for a system size $N = 128$.  
The visualisations of this process at selected time steps of one run
are shown in Fig. \ref{fig:noshearvis}. We observe the development
of the usual sponge microemulsion phase \cite{bib:bce,bib:ecb} consisting
of tubular-like domains; the phase is isotropic in nature.    

\begin{figure}
\begin{center}
\leavevmode
\hbox{%
\epsfxsize=5.5in
\epsffile{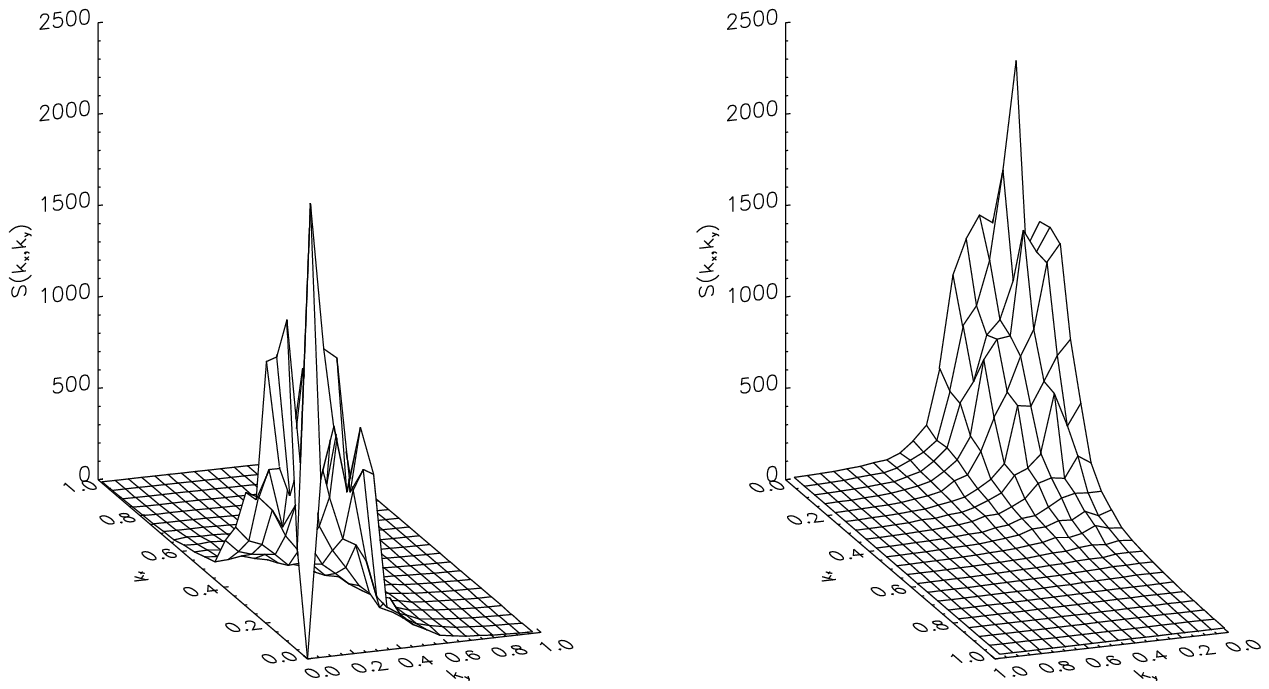}}
\end{center}
\caption{\sl Time-averaged structure factor $S(\bfk, \tau)$ for sponge
microemulsion case with no shear velocity. The system size is $128
\times 128$. Values of $\tau$ depicted here are
$8,000$ time steps on the left-hand side and $12,000$ time steps on the
right. The axes on the
right-hand figure are a mirror image of those on the left to aid
visual clarity.}  
\label{fig:noshearsf1}
\end{figure}

\begin{figure}
\begin{center}
\leavevmode
\hbox{%
\epsfxsize=5.5in
\epsffile{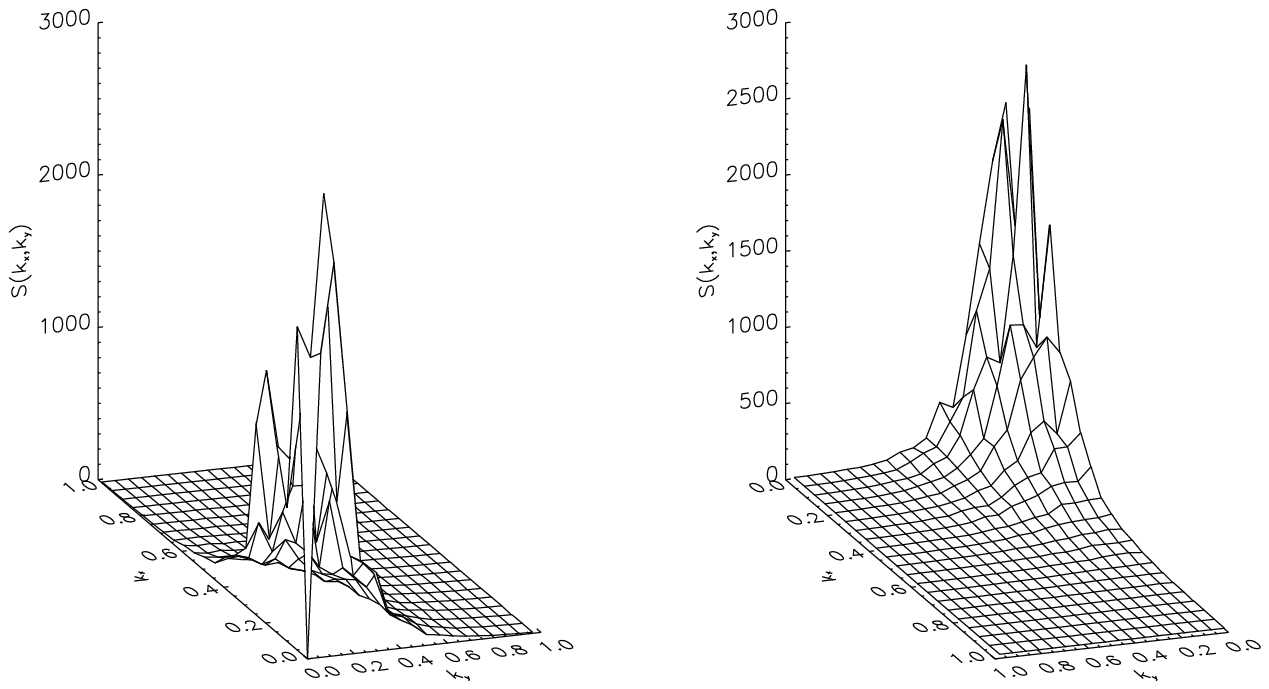}}
\end{center}
\caption{\sl Time-averaged structure factor $S(\bfk, \tau)$ for sponge
microemulsion case with no shear velocity. The system size is $128
\times 128$. Values of $\tau$ depicted here are
$28,000$ time steps on the left-hand side and $36,000$ time steps on the
right. The axes on the
right-hand figure are a mirror image of those on the left to aid
visual clarity.}  
\label{fig:noshearsf2}
\end{figure}

We have calculated the time-averaged
structure factor of the oil-water density every $4,000$ time steps 
during one run, which we then ensemble-averaged over five independent
runs. For typical results at different values of $\tau$ see
Figs \ref{fig:noshearsf1} \& \ref{fig:noshearsf2}. The sponge phase is
isotropic; however, the system has a preferred length scale. This
length scale corresponds to a finite wave length, but not to a specific
wave vector in the structure function. It is clear from
our plots that the growth of structure has no preference for any
lattice direction: various peaks at non-zero wave vectors and a high
background intensity around a wavelength $\left|\bf{k}\right| \approx
0.11$ characterize our system. When we
analyze the structure factor more carefully we find that its maximal
peak is not at a constant wave vector, but at a constant wave length,
as expected in the isotropic sponge phase. 

Additionally, as we found in our previous work
\cite{bib:bce}, the domain structures do not grow in size indefinitely.
Rather, what we see is the development of an equilibrium sponge phase;
however, we note that the underlying lattice dynamics are still
present. The characteristic size of the oil-water domains has 
stopped growing before time step $8,000$. The peak height in the
structure factor has then reached a level of approximately $3,000$ and
stays at this value for the rest of the simulation. Comparative runs
with a system size $N = 64$ showed no difference in this behaviour.
  
The other measurement
we make is of the sums $X^2$ and $Y^2$ of the squared $x$ and $y$
components of all surfactant vectors in the system at every time step,
eqns~(\ref{X}) and~(\ref{Y});
again these values are calculated for later comparison with the case
when shear is imposed. The data at
various time steps are shown in Table I. In this
no-shear case, as
expected, there is essentially no difference between the $X^2$ and $Y^2$
components during the time scale of the simulation, this being further
evidence for the presence of an isotropic system.  

In conclusion, from our results we can say that without shear flow our
system quickly reaches the isotropic sponge equilibrium phase. We will
now turn our attention to the case when shear flow is present, with all the
other parameters in the system remaining fixed.
  
\subsection{Application of Shear}
\label{sec:shear}

\begin{figure}
\begin{center}
\leavevmode
\hbox{%
\epsfxsize=5.5in
\epsffile{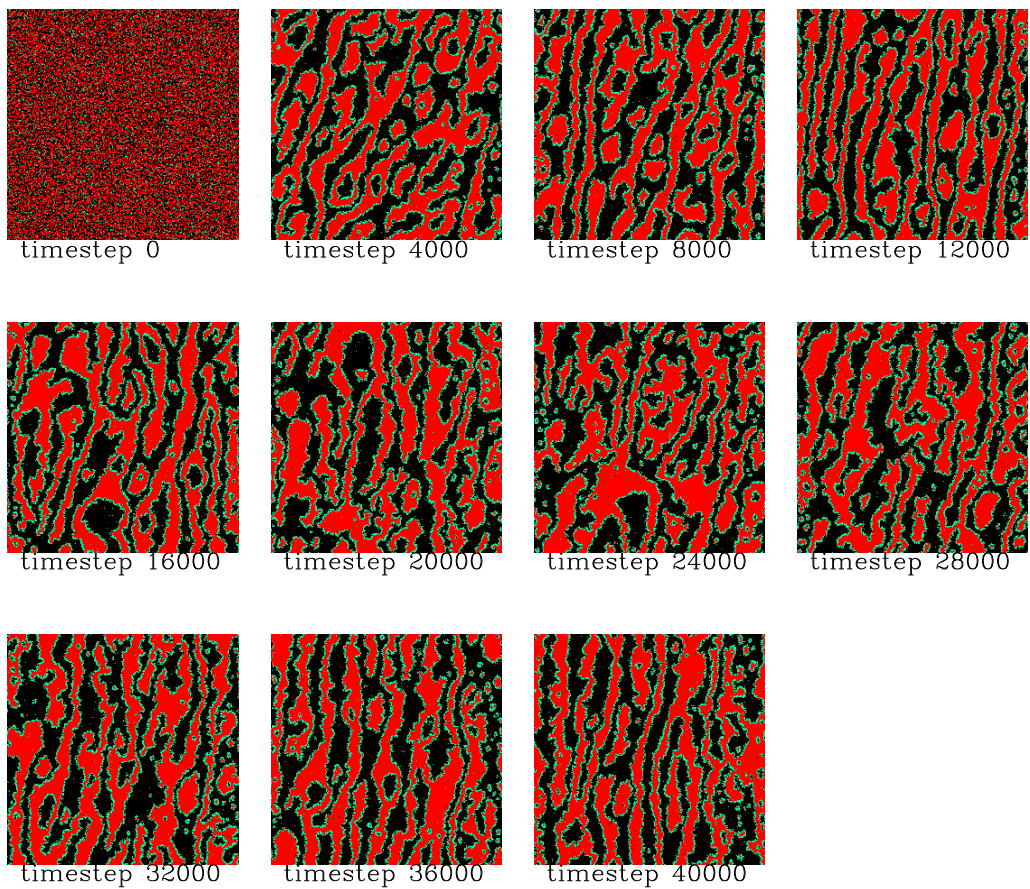}}
\end{center}
\caption{\sl Time evolution of lamellar phase for a system size $256
\times 256$ in the presence
of shear. The $x$ axis is along the horizontal and the $y$ 
axis along the vertical side of the simulation images shown here. The 
imposed shear velocity is $\pm 0.2$ lattice units per time step in the 
$y$ direction, $+0.2$ being on the right.}
\label{fig:shearvis}
\end{figure}

We study the effect of shear flow on our system for three different
lattice sizes, with $N = 64, 128$ and $256$. As discussed in
Sec. \ref{sec:r}, we have to adjust the applied shear velocity in the
$y$ direction of the simulation box accordingly. We have performed a
minimum of five statistically independent runs for each of these
lattice sizes in order to obtain good statistics and consistency of
our results. 
A typical visual result we obtain for a system size of $256 \times
256$ is shown in Fig. \ref{fig:shearvis}. The difference from the case
without shear is dramatic: the shear causes the system to form
lamellar-like objects, which finally connect by wrapping around the
simulation box and therefore -- due to the periodic boundary
conditions -- extend `infinitely' in the $y$ direction. The lamellae are
formed correctly with oil and water rich layers separated by a thin layer
of surfactant and are oriented perpendicular to the velocity gradient
as found in experiments in hyperswollen lyotropic systems
\cite{bib:yt}. There is a clear orientational ordering in the system
and -- since the lamellae are of equal width -- also
evidence for positional ordering.  
It is also obvious
from the visualisation that the underlying dynamics of our model cause
long transient times until the lamellar phase has stabilised.  

\begin{figure}
\begin{center}
\leavevmode
\hbox{%
\epsfxsize=5.5in
\epsffile{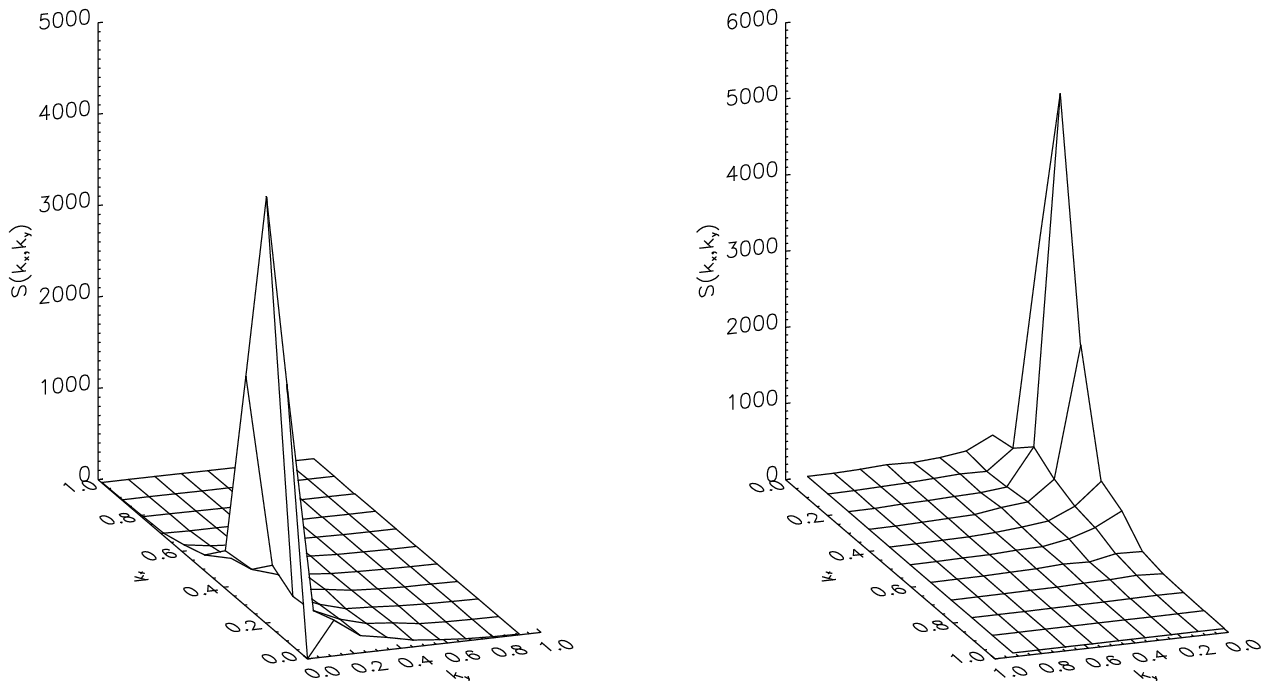}}
\end{center}
\caption{\sl Time-averaged structure factor $S(\bfk, \tau)$ for lamellar
case with shear. The system size is $64 \times 64$ and the shear
velocity is $\pm 0.05$. Values of $\tau$ depicted here are
$28,000$ time steps on the left-hand side and $36,000$ time steps on the
right. The axes on the right-hand figure are a mirror image of those
on the left to aid visual clarity.}  
\label{fig:64shearsf}
\end{figure}

\begin{figure}
\begin{center}
\leavevmode
\hbox{%
\epsfxsize=5.5in
\epsffile{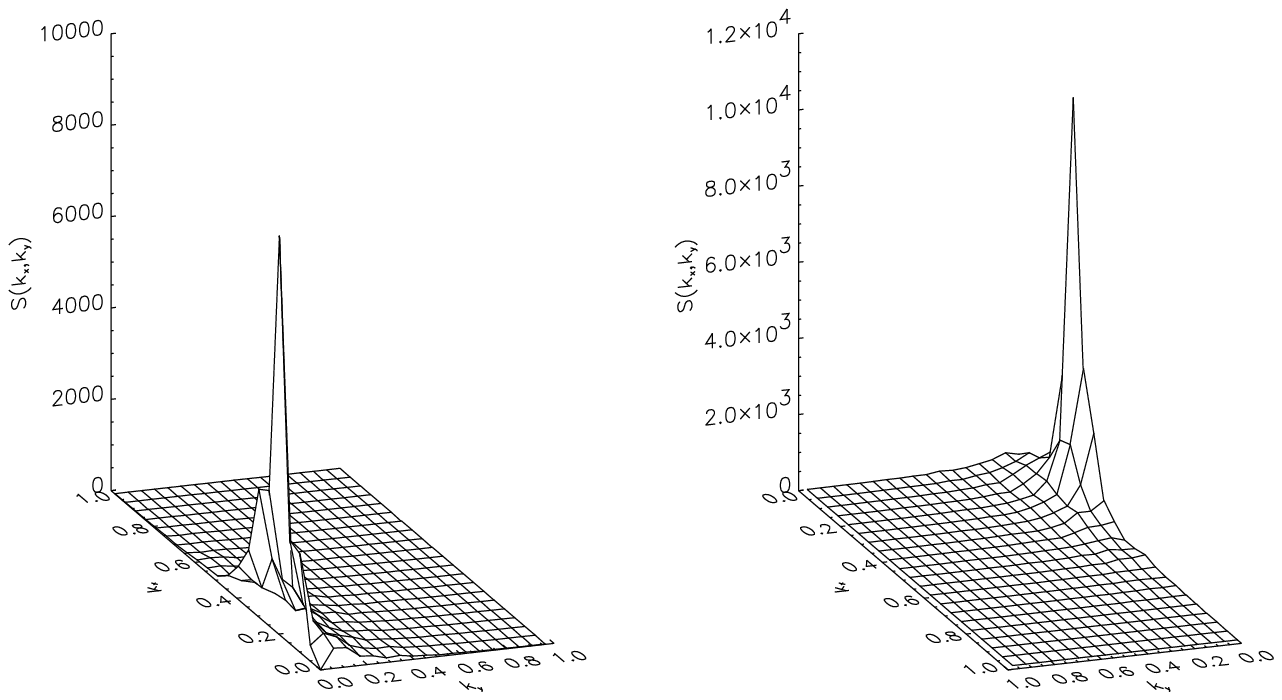}}
\end{center}
\caption{\sl Time-averaged structure factor $S(\bfk, \tau)$ for lamellar
case with shear. The system size is $128 \times 128$ and the shear
velocity is $\pm 0.1$. Values of $\tau$ depicted here are
$48,000$ time steps on the left-hand side and $56,000$ time steps on the
right. The axes on the right-hand figure are a mirror image of those
on the left to aid visual clarity.}  
\label{fig:128shearsf}
\end{figure}

\begin{figure}
\begin{center}
\leavevmode
\hbox{%
\epsfxsize=5.5in
\epsffile{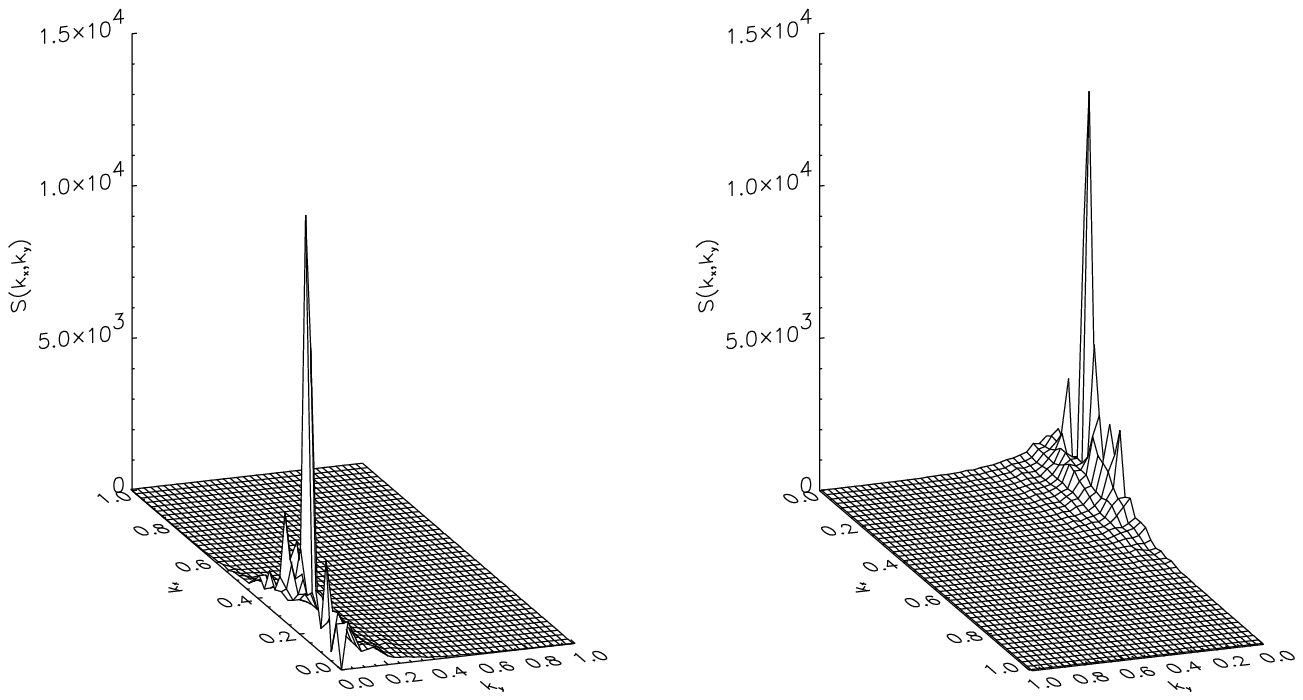}}
\end{center}
\caption{\sl Time-averaged structure factor $S(\bfk, \tau)$ for lamellar
case with shear. The system size is $256 \times 256$ and the shear
velocity is $\pm 0.2$. Values of $\tau$ depicted here are
$40,000$ time steps on the left-hand side and $48,000$ time steps on the
right. The axes on the right-hand figure are a mirror image of those
on the left to aid visual clarity.}  
\label{fig:256shearsf}
\end{figure}

To gain more evidence for the formation of a lamellar phase in our
system we also repeated the calculation of the 
structure factor of the oil-water density. Typical results are shown
in Fig. \ref{fig:64shearsf} for $N = 64$, in Fig. \ref{fig:128shearsf}
for $N = 128$ and in Fig. \ref{fig:256shearsf} for $N = 256$. 
Again, we obtain very different behaviour from the case of no
shear. In all simulations we observe the formation of a clear peak in
the structure factor at $k_y = 0.0$ and $k_x \approx 0.18$, indicating
structures that are infinitely extended in the $y$ direction and
periodic in the $x$ direction. The periodic ordering in the $x$
direction appears to be sinusoidal rather than a square wave. We
believe that the presence of surfactant, which carries colour charge
(order parameter) $q = 0$, smoothes the ordering and hence suppresses
the higher harmonics. The height of
this peak is essentially constant after $32,000$ time steps,
indicating that our system is now in an equilibrium state. In most
simulations we observe another one or even two more peaks at earlier time
steps, all with $k_y = 0.0$ but different $k_x$.\footnote{These peaks
are not harmonics of the first peak, since they do not appear at
integer multiples of the first.} We believe that these
peaks originate from competing lamellar widths in our system and
therefore cause long transient effects in our simulations. We
have, however, run several of our simulations for over $60,000$ time
steps and the peak height and position remained stable in these
simulations. 

\begin{figure}
\begin{center}
\leavevmode
\hbox{%
\epsfxsize=4.0in
\epsffile{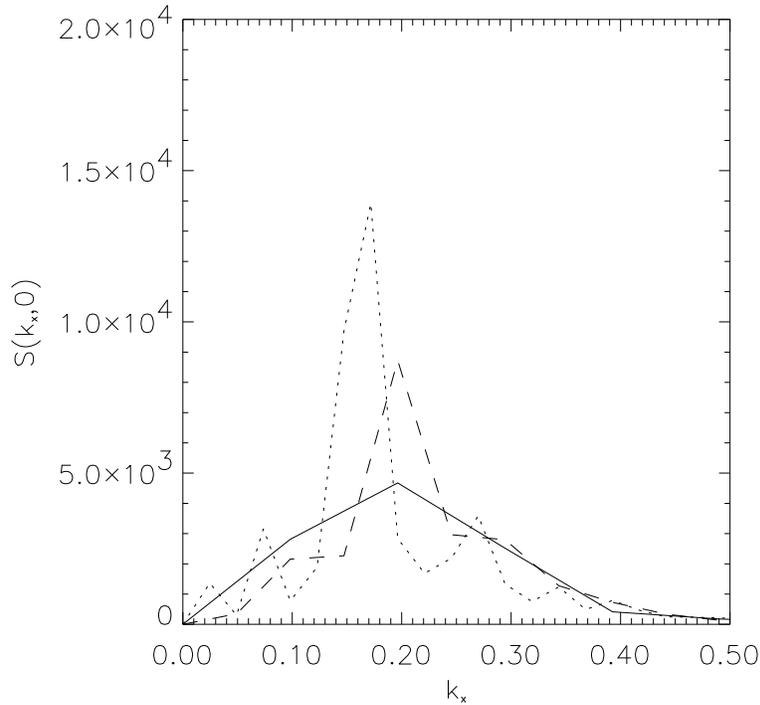}}
\end{center}
\caption{\sl Time averaged structure factor at $k_y=0.0$ for system
sizes $N = 64$ (solid line), $N = 128$ (dashed line) and $N = 256$
(dotted line) plotted against $k_x$. The data is taken from the
left-hand plots of Figures $6, 7 \& 8$ respectively. Note that the
shift in the position of the peak is due to the discretisation of the
wave vectors (compare discussion in Sec. III.B.)}  
\label{fig:peaksize}
\end{figure}

The existence of a single peak in $S(\bfk)$, however, is not complete
evidence for lamellar ordering \cite{bib:c}. Hence, we additionally
studied the
behaviour of the peak as a function of system size. For a truly
ordered state, one expects sharpening and divergence of the peak when
the system size is increased. In Fig. \ref{fig:peaksize}, where we
have plotted $S(k_x)$ at $k_y = 0.0$ for the peaks from
Figs. \ref{fig:64shearsf}, \ref{fig:128shearsf} $\&$
\ref{fig:256shearsf}, this behaviour is obvious. The shift in the
position of the peak at system size $N = 256$ is due to the fact that,
owing to the discretisation of $k_x$ the value of $k_x = 0.1718$ is not
available for $N = 64$ or $N = 128$. The peak instead appears at the
nearest wavevector, being $k_x = 0.1963$. From this result we can
conclude that in the case of shear flow, our system is truly in an
ordered phase.  

Finally, we also looked at the sums $X^2$ and $Y^2$ of the squared $x$
and $y$ components of all surfactant vectors. The values of
these expressions at selected time steps for system size $N = 128$
are shown in Table II. 
As expected, the surfactant vectors starting from an isotropic
distribution at time step $t = 0$ align perpendicular to the direction of
shear, hence $X^2 > Y^2$. This indicates again the formation of
lamellae which are extended in the $y$ direction. The corresponding
values for $N = 64$ and $N = 256$ give similar results; the ratio
$X^2 : Y^2$ increases slightly with system size.  

Summing up our evidence, we have established that under the influence
of shear our system no longer evolves to the isotropic sponge
microemulsion phase but instead to the lamellar phase with structures
which are periodic in one dimension and infinitely extended in the other. 
     
\begin{figure}
\begin{center}
\leavevmode
\hbox{%
\epsfxsize=5.5in
\epsffile{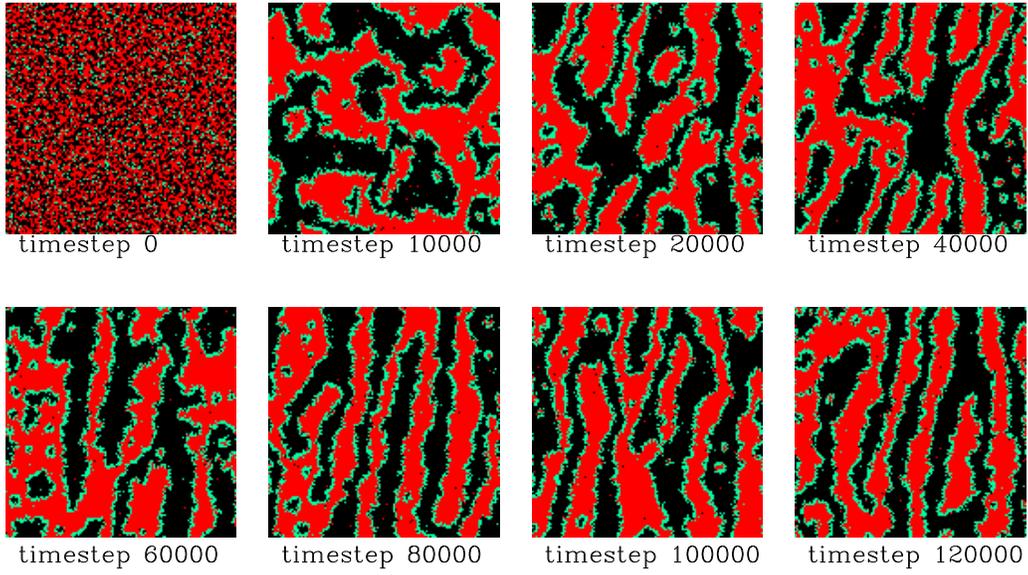}}
\end{center}
\caption{\sl Time evolution of ternary amphiphilic system, shear imposed only
after time step $10,000$. The $x$ axis is along the horizontal and the $y$ 
axis along the vertical side of the simulation images shown here. The 
imposed shear velocity is $\pm 0.1$ lattice units per time step in the 
$y$ direction, $+0.1$ being on the right. The system size is $128
\times 128$.}
\label{fig:longshearvis}
\end{figure}

\begin{figure}
\begin{center}
\leavevmode
\hbox{%
\epsfxsize=5.5in
\epsffile{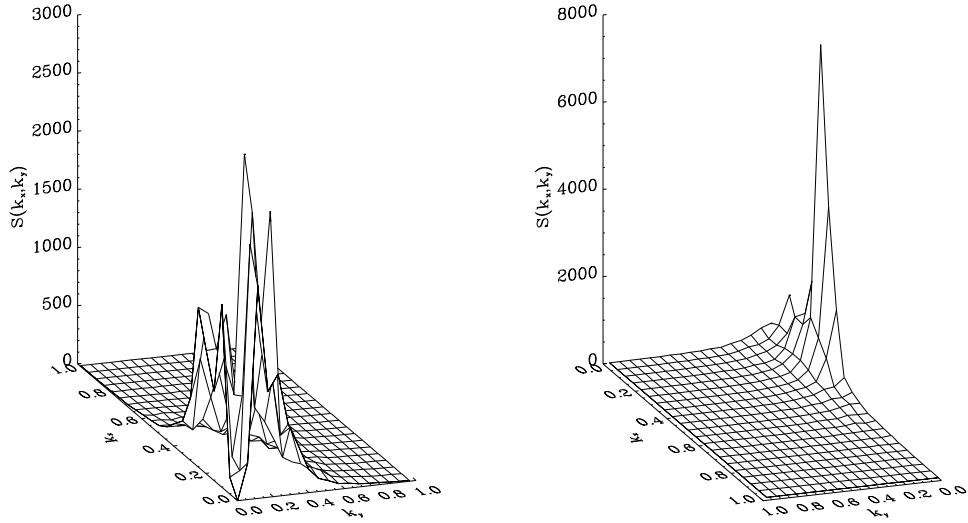}}
\end{center}
\caption{\sl Time-averaged structure factor $S(\bfk, \tau)$ for the
case with shear imposed only after $10,000$ time steps. The system
size is $128 \times 128$ and the shear
velocity is $\pm 0.1$. Values of $\tau$ depicted here are
$10,000$ time steps on the left-hand side and $120,000$ time steps on the
right. The axes on the right-hand figure are a mirror image of those
on the left to aid visual clarity.}  
\label{fig:longshearsf}
\end{figure}

In addition, we have confirmed that the lamellar state is not simply a
shear-induced pattern of phase separated fluid domains, but actually
results from perturbing the equilibrium sponge microemulsion phase. This is
accomplished by imposing shear only after time step $10,000$ has been
reached in a simulation of lattice size $128 \times 128$, by which
point the domain structure formed is that of a sponge equilibrium
phase \cite{bib:bce,bib:ecb}; this then
comes under the influence of the shear flow. Although we now have to  
wait longer for the phase transition to occur, the system still evolves from
the isotropic, equilibrium phase to the anisotropic lamellar state as
we would expect (see Figs. \ref{fig:longshearvis} $\&$
\ref{fig:longshearsf}). Both the visualisation and the structure
factor show the characteristics of the sponge phase at $t =
10,000$, but following time step $t = 80,000$ the system is in the
lamellar phase.

\section{Conclusions}
\label{sec:conc}

On general theoretical grounds, one does not expect an isotropic-to-smectic 
transition at equilibrium at nonzero temperature in two dimensions. However, 
using our hydrodynamic lattice-gas model, we 
have been able to simulate the transition from an isotropic
sponge to a lamellar phase in a two dimensional ternary amphiphilic system
under the influence of an applied linear shear. This finding implies that
shear shifts the isotropic-to-smectic transition point from 
zero to nonzero temperature. Our work confirms that the model
is capable of describing complex multi-phase fluid phenomena that are
currently out of reach of other simulation methods. To link our results
more fully with both experimental data and theoretical analysis,
a three dimensional version of the present model \cite{bib:bc} and a
detailed investigation of the complete non-equilibrium phase diagram
are required. These represent an area of ongoing and future work. 

\section*{Acknowledgments}

We are grateful to Mike Cates and John Olson for helpful
discussions during the development of this work. In particular, we
thank John Olson for assistance in implementing his shear 
methods within our model. ANE wishes to thank EPSRC and
Schlumberger for funding his CASE award. FWJW is indebted to the
Stiftung Maximilianeum, M\"{u}nchen, and Balliol College, Oxford
University, for supporting his stay in Oxford. PVC is grateful to
Wolfson College and the Department of Theoretical Physics, Oxford
University, for a Visiting Fellowship (1996-1998). BMB was
supported in part by Phillips Laboratories and by the United States
Air Force Office of Scientific Research under grant number
F49620-95-1-0285. PVC and BMB thank NATO grant number CRG950356 and
the CCP5 committee of EPSRC for funding a visit to U.K. by BMB.

\newpage

\newpage

\newpage

\begin{table}[bh] 
\label{tab:noshear}
\vspace{0.1cm}
\centering
{
\renewcommand{\baselinestretch}{1}
\small\normalsize
\vspace{0.5cm}
\begin{tabular}{rlll}
\hline
\em{Time Step}  &  \em{$X^2$} &  \em{$Y^2$} & \em{$X^2$}:\em{$Y^2$} \\
\hline
    0 & 6148$\pm$  68 & 6147$\pm$  52 &  1.000$\pm$ 0.014 \\ 
 4000 & 6081$\pm$ 144 & 6214$\pm$ 166 &  0.979$\pm$ 0.035  \\
 8000 & 6046$\pm$  33 & 6249$\pm$  91 &  0.968$\pm$ 0.015  \\
12000 & 6161$\pm$ 137 & 6134$\pm$ 143 &  1.005$\pm$ 0.033 \\
16000 & 6129$\pm$ 160 & 6166$\pm$ 152 &  0.994$\pm$ 0.036 \\
20000 & 6169$\pm$ 103 & 6126$\pm$  92 &  1.007$\pm$ 0.023 \\
24000 & 6148$\pm$ 144 & 6147$\pm$ 169 &  1.000$\pm$ 0.036 \\
28000 & 6135$\pm$ 150 & 6160$\pm$ 155 &  0.996$\pm$ 0.035 \\
32000 & 6150$\pm$ 101 & 6145$\pm$ 103 &  1.001$\pm$ 0.024 \\
36000 & 6113$\pm$  56 & 6182$\pm$  93 &  0.989$\pm$ 0.017 \\
40000 & 6107$\pm$ 147 & 6188$\pm$ 155 &  0.987$\pm$ 0.034 \\
\hline
\end{tabular}
\caption{\sl Total $X$ and $Y$ component values of surfactant dipole
vectors at selected time steps for the case of no shear and $N = 128$. The ratio 
{\em{$X^2$}:\em{$Y^2$}} is also shown. The error bars result from
ensemble averaging over five independent runs.} }
\end{table}

\vspace{3cm}

\begin{table}[bh] 
\vspace{0.1cm}
\centering
{
\renewcommand{\baselinestretch}{1}
\small\normalsize
\vspace{0.5cm}
\begin{tabular}{rlll}
\hline
\em{Time Step}  &  \em{$X^2$} &  \em{$Y^2$} & \em{$X^2$}:\em{$Y^2$} \\
\hline
    0 & 6163$\pm$  77 & 6152$\pm$  51 &   1.001$\pm$  0.015 \\
 4000 & 7248$\pm$ 276 & 5068$\pm$ 234 &   1.430$\pm$  0.086 \\
 8000 & 7647$\pm$ 218 & 4669$\pm$ 229 &   1.638$\pm$   0.093 \\
12000 & 7320$\pm$ 136 & 4996$\pm$ 168 &   1.465$\pm$   0.056 \\
16000 & 7348$\pm$ 184 & 4968$\pm$ 238 &   1.479$\pm$   0.080 \\
20000 & 7120$\pm$ 253 & 5196$\pm$ 319 &   1.370$\pm$   0.097 \\
24000 & 7120$\pm$ 193 & 5196$\pm$ 221 &   1.370$\pm$   0.069\\
28000 & 7083$\pm$ 195 & 5232$\pm$ 225 &   1.354$\pm$   0.069 \\
32000 & 7310$\pm$ 100 & 5006$\pm$ 149 &   1.460$\pm$   0.048 \\
36000 & 7060$\pm$ 125 & 5256$\pm$ 200 &   1.343$\pm$   0.056 \\
40000 & 7211$\pm$ 132 & 5105$\pm$ 219 &   1.412$\pm$   0.066 \\
\hline
\end{tabular}
\caption{\sl Total $X$ and $Y$ component values of surfactant dipole
vectors at selected time steps for the case of shear flow and $N = 128$. The ratio 
{\em{$X^2$}:\em{$Y^2$}} is also shown. The error bars are obtained
from averaging over five independent runs.} }
\label{tab:shear}
\end{table}

\end{document}